\documentclass{llncs}
\usepackage{graphics}
\usepackage{latexsym}

\begin{document}
\title{Scalable Unix Commands for Parallel Processors:  A High-Performance Implementation\thanks{This
work was supported by the Mathematical, Information, and
Computational Sciences Division subprogram of the Office of
Advanced Scientific Computing Research, U.S.\ Department of Energy,
under Contract W-31-109-Eng-38.}}
\author{Emil Ong   \and
        Ewing Lusk \and
        William Gropp }
\institute{Mathematics and Computer Science Division\\
Argonne National Laboratory\\Argonne, Illinois 60439 USA}
\maketitle

\begin{abstract}

We describe a family of MPI applications we call the Parallel Unix Commands.
These commands are natural parallel versions of common Unix user commands such
as {\tt ls}, {\tt ps}, and {\tt find}, together with a few similar commands
particular to the parallel environment.  We describe the design and
implementation of these programs and present some performance results on a
256-node Linux cluster.  The Parallel Unix Commands are open source and freely
available.

\end{abstract}

\section{Introduction}
\label{sec:intro}

The oldest Unix commands ({\tt ls}, {\tt ps}, {\tt find}, {\tt grep}, etc.)
are built into the fingers of experienced Unix users.  Their usefulness has
endured in the age of the GUI not only because of their simple,
straightforward design but also because of the way they work together.  Nearly all
of them do I/O through {\tt stdin} and {\tt stdout}, which can be redirected
from/to files or through pipes to other commands.  Input and output are lines
of text, facilitating interaction among the commands in a way that would be
impossible if these commands were GUI based.

In this paper we describe an extension of this set of tools into the parallel
environment.  Many parallel environments, such as Beowulf clusters and networks
of workstations, consist of a collection of individual machines, with at least
partially distinct file systems, on which these commands are supported.  A
user may, however, want to consider the collection of machines as a single
parallel computer, and yet still use these commands.  Unfortunately, many
common tasks, such as listing files in a directory or processes running on
each machine, can take unacceptably long times in the parallel environment
if performed sequentially, and can produce an inconveniently large amount of
output.  

A preliminary version of the specification of our Parallel Unix Commands appeared
in~\cite{sut}.  New in this paper are a refinement of the specification based
on experience, a high-performance implementation based on MPI for improved
scalability, and measurements of performance on a 256-node Unix cluster.

The tools described here might be useful in the construction of a
cluster-management system, but this collection of user commands does not
itself purport to {\em be\/} a cluster-management system, which needs more
specialized commands and a more extensive degree of fault tolerance.  Although
nothing prevents these commands from being run by {\tt root} or being integrated
into cluster-management scripts, their primary anticipated use is the same as
that of the classic Unix commands:  interactive use by ordinary users to carry
out their ordinary tasks.

\section{Design}
\label{sec:design}

In this section we describe the general principles behind this design and then
the specification of the tools in detail.

\subsection{Goals}
\label{sec:goals}

The goals for this set of tools are threefold:

\begin{itemize}
\item They should be familiar to Unix users.  They should have
  easy-to-remember names (we chose {\tt pt<unix-command-name>}) and take the
  same arguments as their traditional counterparts to the extent consistent
  with the other goals.
\item They should interact well with other Unix tools by producing output that
  can be piped to other commands for further processing, facilitating the
  construction of specialized commands on the command line in the classic Unix
  tradition. 
\item They should run at interactive speeds, as do traditional Unix commands.
  Parallel process managers now exist that can start MPI programs quickly,
  offering the same experience of immediate interaction with the parallel
  machine, while providing information from numerous individual machines.
\end{itemize}

\subsection{Specifying Hosts}
\label{sec:hosts}

All the commands use the same approach to specifying the collection of hosts
on which the given command is to run.  A host list can be given either
explicitly, as in the blank-separated list {\tt 'donner dasher blitzen'}, or
implicitly in the form of a pattern like {\tt ccn\%d@1-32,42,65-96}, which
represents the list {\tt ccn1,...,ccn32,ccn42,ccn65,...,ccn96}.

All of the commands described below have a hosts argument as an (optional)
first argument.  If the environment variable {\tt PT\_MACHINE\_FILE} is set,
then the list of hosts is read from the file named by the value of that
variable.  Otherwise the first argument of a command is one of the following:
\begin{description}
\item[{\tt -all}] all of the hosts on which the user is allowed to run,
\item[{\tt -m}] the following argument is the name of a file containing the
  host names,
\item[{\tt -M}] the following argument is an explicit or pattern-based list of machines.
\end{description}
Thus 
\begin{verbatim}
    ptls -M "ccn%d-myr@129-256" -t /tmp/lusk
\end{verbatim}
runs a parallel version of {\tt ls -t} (see below) on the directory {\tt /tmp/lusk}
on nodes with names {\tt ccn129-myr,\ldots,ccn256-myr}.

\subsection{The Commands}
\label{sec:commands}

The Parallel Unix Commands are shown in Table~\ref{tab:commands}.
They are of three types:  straightforward parallel versions of traditional
commands with little or no output;  parallel versions of traditional commands
with specially formatted output; and new commands in the spirit of the
traditional commands but particularly inspired by the parallel environment.
\begin{table}[ht!]
\begin{center}
\caption{Parallel UNIX Commands}
\begin{tabular}{l|l}
\textbf{Command}&\textbf{Description}\\
\hline
\texttt{ptchgrp}&Parallel \texttt{chgrp}\\
\texttt{ptchmod}&Parallel \texttt{chmod}\\
\texttt{ptchown}&Parallel \texttt{chown}\\
\texttt{ptcp}&Parallel \texttt{cp}\\
\texttt{ptkillall}&Parallel \texttt{killall}\\
&(Linux semantics)\\
\texttt{ptln}&Parallel \texttt{ln}\\
\texttt{ptmv}&Parallel \texttt{mv}\\
\texttt{ptmkdir}&Parallel \texttt{mkdir}\\
\texttt{ptrm}&Parallel \texttt{rm}\\
\texttt{ptrmdir}&Parallel \texttt{rmdir}\\
\texttt{pttest[ao]}&Parallel \texttt{test}\\
\end{tabular}
\hspace{1cm}
\begin{tabular}{l|l}
\textbf{Command}&\textbf{Description}\\
\hline
\texttt{ptcat}&Parallel \texttt{cat}\\
\texttt{ptfind}&Parallel \texttt{find}\\
\texttt{ptls}&Parallel \texttt{ls}\\
\hline
\texttt{ptfps}&Parallel process\\
&\hspace{.2cm}space find\\
\texttt{ptdistrib}&Distribute files\\
&\hspace{.2cm}to parallel jobs\\
\texttt{ptexec}&Execute jobs in\\
&\hspace{.2cm}parallel\\
\texttt{ptpred}&Parallel predicate\\
\end{tabular}
\vspace{.25cm}
\label{tab:commands}
\end{center}
\end{table}

\vspace{-1.5cm}
\subsubsection{Parallel Versions of Traditional Commands}
\label{sec:traditional}

The first part of Table~\ref{tab:commands} lists the commands that are simply
common Unix commands that are to be run on each host.
The semantics for many of these is very natural -- the corresponding
uniprocessor version of any command is run on every node specified.  For
example, the command
\begin{verbatim}
    ptrm -M "node%d@1-5" -rf old_files/
\end{verbatim}
is equivalent to running
\begin{verbatim}
    rm -rf old_files/
\end{verbatim}
on node1, node2, node3, node4, and node5.  The command line arguments to most
of the commands have the same meaning as their uniprocessor counterparts.

The exceptions \texttt{ptcp} and \texttt{ptmv} deserve special mention;
the semantics of parallel copy and move are not necessarily obvious.  
The commands presented here perform one-to-many copies by using MPI and 
compression;  \texttt{ptmv} deletes the local files that were copied if the 
copy was successful.  The command line arguments for \texttt{ptcp} and 
\texttt{ptmv} are identical to their uniprocessor counterparts with the 
exception of an option flag, \texttt{-o}.  This flag allows the user to 
specify whether compression is used in the transfer of data.  In 
the future the flags may be expanded to allow for other customizations.
Handling of directories as either source or destination is handled as in the
normal version of {\tt cp} or {\tt mv}.

Parallel \texttt{test} also deserves explanation.  There are two
versions of parallel \texttt{test}; both run \texttt{test} on all specified
nodes, but \texttt{pttesta} logically ANDs the results of the tests,
while \texttt{pttesto} logically ORs the results of the tests.  By
default, \texttt{pttest} is an alias for \texttt{pttesto}.  This link
allows the natural semantics of \texttt{pttest} to detect failure on any
node.

\subsubsection{Parallel Versions of Common UNIX Commands with Formatted Output}
\label{sec:formatted}

The second set of commands in Table~\ref{tab:commands} may
produce a significant amount of output.  In order to facilitate handling
of this output, if the first argument to {\tt ptfind}, {\tt ptls}, or {\tt
  ptcat} is {\tt -h} (for ``headers''), then the output from each host will be
preceded by a line identifying the host.  This is useful for piping into other
commands such as {\tt ptdisp} (see below).  In the example
\begin{verbatim}
    $ ptls -M "node%d@1-3" -h
    [node1.domain.tld]
    myfile1
    [node2.domain.tld]
    [node3.domain.tld]
    myfile1
    myfile2
\end{verbatim}
the user has file \texttt{myfile1} on node1, no files in the
current directory on node2, and the files \texttt{myfile1} and
\texttt{myfile2} on node3.  All other command line arguments to these commands
have the same meaning as their uniprocessor counterparts.

To facilitate processing later in a pipeline by filters such as {\tt grep}, we
provide a filter that {\em spreads\/} the hostname across the lines of
output, that is,
\begin{verbatim}
    $ ptls -M "node%d@1-3" -h | ptspread
    node1.domain.tld:  myfile1
    node3.domain.tld:  myfile1
    node3.domain.tld:  myfile2
\end{verbatim}

\subsubsection{New parallel commands}

The third part of Table~\ref{tab:commands} lists commands that are in the
spirit of the other commands but have no non-parallel counterpart.

Many of the uses of {\tt ps} are similar to the uses of {\tt ls}, such as
determining the age of a process (respectively, a file) or owner of a process
(respectively, a file).  Since a Unix file system typically contains a large
number of files, the Unix command {\tt find}, with its famously awkward
syntax, provides a way to search the file system for files with certain
combinations of properties.  On a single system, there are typically not so
many processes running that they cannot be perused with {\tt ps} piped to {\tt
  grep}, but on a parallel system with even a moderate number of hosts, a {\tt
  ptps} could produce thousands of lines of output.  Therefore, we have
proposed and implemented a counterpart to {\tt find}, called {\tt ptfps}, that
searches the process space instead of the file space.  In the Unix tradition
we retain the syntax of {\tt find}.  Thus
\begin{verbatim}
    ptfps -all -user lusk
\end{verbatim}
will list all the processes belonging to user {\tt lusk} on all the machines
in a format similar to the output of {\tt ps}, and 
\begin{verbatim}
     ptfps -all -user gropp -time 3600 -cmd ^mpd
\end{verbatim}
will list all processes owned by {\tt gropp}, executing a command beginning
with {\tt mpd}, that have been running for more than an hour.  Many more
filtering specifications and output formats are available; see the (long) {\tt
  man} page for {\tt ptfps} for details.

The command \texttt{ptdistrib} is effectively a scheduler for running a
command on a set of files over specified nodes.  For example, to compile all
of the C files in the current directory over all nodes currently available,
then fetch back all the resulting files, the user might use the following
command:
\begin{verbatim}
    ptdistrib -all -f 'cc -c {}' *.c
\end{verbatim}
Here, the \texttt{\{\}} is replaced by the names of the files given, one by
one.  See the {\tt man} page for more information.

The command \texttt{ptexec} simply executes a command on all nodes. To
determine, for example, which hosts were available for running
jobs, the user might run the following command:
\begin{verbatim}
    ptexec -all hostname
\end{verbatim}
No special formatting of output or return code checking is done.

The command \texttt{ptpred} runs a \texttt{test} on each specified node and
outputs a 0 or 1 based on the result of the test.  For example, to test for
the existence of the file {\tt myfile} on nodes node1, node2, and node3, the
user might have the following session:

\begin{verbatim}
    $ ptpred -M "node1 node2 node3" '-f myfile'
    node1.domain.tld: 1
    node2.domain.tld: 0
    node3.domain.tld: 1
\end{verbatim}

\noindent
In this case, node1 and node3 have the file, but node2 does not.  Note that
\texttt{ptpred} prints the logical result of \texttt{test}, not the verbatim
return value.

\begin{figure}[ht!]
\begin{center}
\scalebox{.5}{
\includegraphics{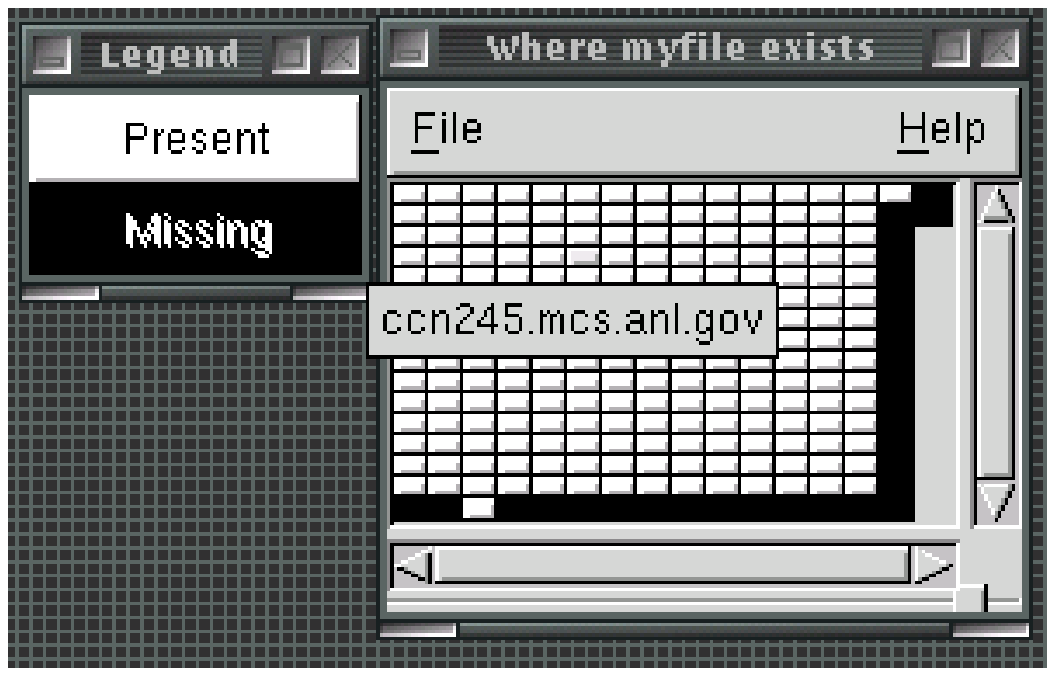}
}
\hspace{1cm}
\scalebox{.5}{
\includegraphics{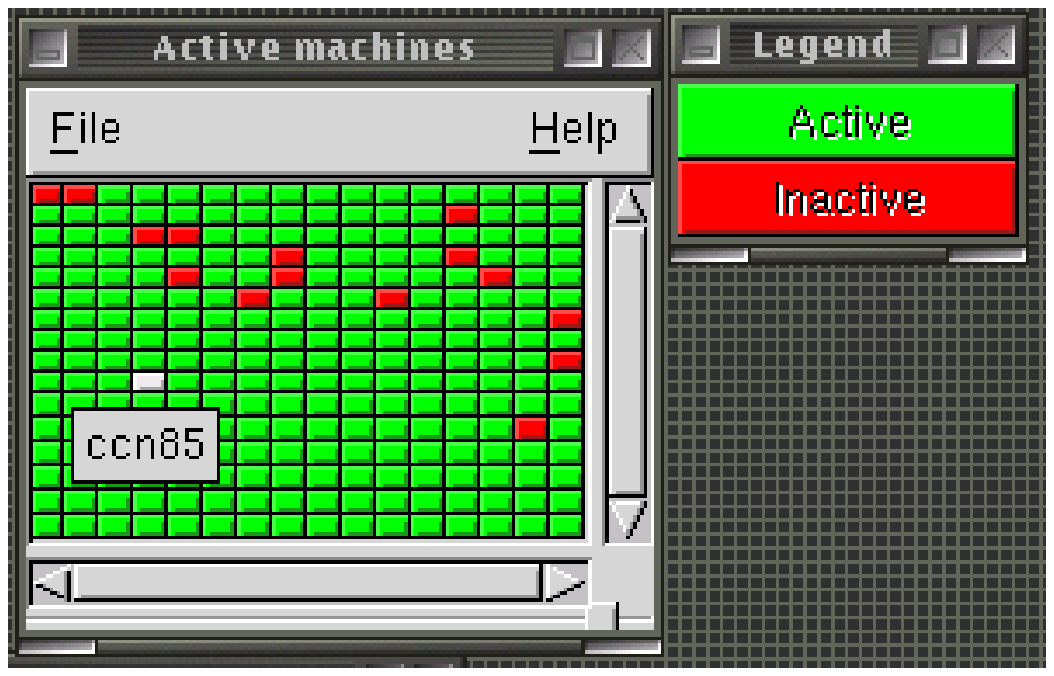}
}
\caption{Screenshots from {\tt ptdisp}}
\label{fig:ptdisp}
\end{center}
\end{figure}

The output of \texttt{ptpred} can be customized:
\begin{verbatim}
    $ ptpred -M "node1 node2 node3" '-f myfile' \
                 'color black green' 'color black red'
    node1.domain.tld: color black green
    node2.domain.tld: color black red 
    node3.domain.tld: color black green
\end{verbatim}
This particular customization is useful as input to \texttt{ptdisp},
which is a  general display tool for
displaying information about large groups of machines.  As an example,
Figure~\ref{fig:ptdisp} shows some screenshots produced by \texttt{ptdisp}.

The command  \texttt{ptdisp} accepts special input from standard input of the form
\begin{verbatim}
    <hostname>: <command> [arguments]
\end{verbatim}
where {\tt command} is one of {\tt color}, {\tt percentage}, {\tt text}, or a
number.  The output corresponding to each host is assigned to one member of an
array of button boxes.

As an example, one might produce the screenshot on the left in
Figure~\ref{fig:ptdisp} with the following command:
\begin{verbatim}
    ptpred -all '-f myfile' 'color black white' \
                            'color white black' \
    | ptdisp -c -t "Where myfile exists"
\end{verbatim}
to find on which nodes a particular file is present.
\vspace{-1cm} The command {\tt ptdisp} can confer scalability on the output of
other commands not part of this tool set by serving as the last step in any
pipeline that prepares lines of input in the form it accepts.  Since it reads
perpetually, it can even serve as a crude graphical system monitor, showing
active machines, as on the right side of Figure~\ref{fig:ptdisp}.  The command
to produce this display is given in Section~\ref{sec:examples}.  The number of
button boxes in the display adapts to the input.  When the cursor is placed
over a box, the node name automatically appears, and clicking on a button box
automatically starts an {\tt xterm} with an {\tt ssh} to that host if
possible, for remote examination.

\section{Examples}
\label{sec:examples}

Here we demonstrate the flexibility of the command set by presenting a few
examples of their use.
\begin{itemize}
\item To look for nonstandard configuration files: 
\begin{verbatim}
 ptcp -all mpd.cfg /tmp/stdconfig; \
 ptexec -all -h diff /etc/mpd.cfg /tmp/stdconfig \
 | ptspread 
\end{verbatim}
This shows differences between a standard file and the version on each node.
\item To look at the load average on the parallel machine:
\begin{verbatim}
 ptexec -all 'echo -n `hostname` ; uptime' | awk '{ print $1 \
 ": percentage " $(NF-1)*25 }' | sed -e 's/,//g' | ptdisp
\end{verbatim}
The {\tt percentage} command to {\tt ptdisp} shows color-coded load averages
in a compact form.
\item To continuously monitor the state of the machine (nodes up or down)
\begin{verbatim}
 (echo "$LEGEND$: Active black green Inactive black red"; \
 while true; do (enumnodes -M 'ccn%d@1-256' \
 | awk '{print $1 ": 0"}') ; sh ptping.sh 'ccn%d@1-256'; \
 sleep 5; done) | ptdisp -t "Active machines" -c
\end{verbatim}
  We assume here that {\tt ptping} pings all the nodes.  This is admittedly
  ugly, but it illustrates the power of the Unix command line and the
  interoperability of Unix commands.  The output of this command is what
  appears on the right side of Figure~\ref{fig:ptdisp}.
\item To kill a runaway job
\begin{verbatim}
 ptfps -all -user ong -time 10000 -kill SIGTERM 
\end{verbatim}
\end{itemize}

\section{Implementation}
\label{sec:implementation}

The availability of parallel process managers, such as
MPD~\cite{bgl00:mpd:pvmmpi00}, that provide pre-emption of existing
long-running jobs and fast startup of MPI jobs, has made it possible to write
these commands as MPI application programs.  Each command parses its hostlist
arguments and then starts an MPI program (with {\tt mpirun} or {\tt mpiexec})
on the appropriate set of hosts.  It is assumed that the process manager and
MPI implementation can manage {\tt stdout} from the individual processes in
the same way that MPD does, by routing them to the {\tt stdout} of the {\tt
  mpirun} process. The graphical output of {\tt ptdisp} is provided by GTK+
(See {\tt http://www.gtk.org}).

Using MPI lets us take advantage of the MPI collective
operations for scalability in delivering input arguments and/or data and
collecting results.  Some of the specific uses of MPI collective operations
are as follows.
\begin{itemize}
\item {\tt MPI\_Bcast} uses {\tt ptcp} to move data to the target nodes.
\item {\tt MPI\_Reduce}, with {\tt MPI\_MIN} as the reduction operation, is
  used in many commands for error checking.
\item {\tt MPI\_Reduce}, with {\tt MPI\_LOR} or {\tt MPI\_LAND} as the
  reduction operation, is used in {\tt pttest}.
\item {\tt MPI\_Gather} is used in {\tt ptdistrib} to collect data enabling
  dynamic reconfiguration of the list of nodes work is distributed to.
\item Dynamically-created MPI communicators other than {\tt MPI\_COMM\_WORLD}
  are used when the task is different on different nodes.  An example of this
  situation occurs when the target specified in the {\tt ptcp} command turns
  out to be a file on some nodes and a directory on others.
\end{itemize}
The implementation of {\tt ptcp} is roughly that described
in~\cite{gropp-lusk-thakur:usingmpi2}.  Parallelism is achieved at three
levels:  writing the file to the local file systems on each host is done in
parallel; a scalable implementation of  {\tt MPI\_Bcast} provides parallelism
in the sending of data; and the files are sent in blocks, providing pipeline
parallelism.  We also use compression to reduce the amount of data that must
be transferred over the network.  Directory hierarchies are {\tt tar}red as
they are being sent.

A user may have different user ids on different machines.  Whether these
scalable Unix commands allow for this situation depends on the MPI
implementation with which they are linked.  In the case of
MPICH~\cite{gropp-lusk:mpich-www}, for example, it is possible for a user to
run a single MPI job on a set of machines where the user has different user
ids.

\section{Performance}
\label{sec:performance}
\begin{figure}[ht!]
  \begin{center}
    \leavevmode
    \scalebox{.8}{
\setlength{\unitlength}{0.240900pt}
\ifx\plotpoint\undefined\newsavebox{\plotpoint}\fi
\sbox{\plotpoint}{\rule[-0.200pt]{0.400pt}{0.400pt}}%
\begin{picture}(1500,900)(0,0)
\font\gnuplot=cmr10 at 10pt
\gnuplot
\sbox{\plotpoint}{\rule[-0.200pt]{0.400pt}{0.400pt}}%
\put(141.0,123.0){\rule[-0.200pt]{4.818pt}{0.400pt}}
\put(121,123){\makebox(0,0)[r]{0}}
\put(1419.0,123.0){\rule[-0.200pt]{4.818pt}{0.400pt}}
\put(141.0,307.0){\rule[-0.200pt]{4.818pt}{0.400pt}}
\put(121,307){\makebox(0,0)[r]{5}}
\put(1419.0,307.0){\rule[-0.200pt]{4.818pt}{0.400pt}}
\put(141.0,492.0){\rule[-0.200pt]{4.818pt}{0.400pt}}
\put(121,492){\makebox(0,0)[r]{10}}
\put(1419.0,492.0){\rule[-0.200pt]{4.818pt}{0.400pt}}
\put(141.0,676.0){\rule[-0.200pt]{4.818pt}{0.400pt}}
\put(121,676){\makebox(0,0)[r]{15}}
\put(1419.0,676.0){\rule[-0.200pt]{4.818pt}{0.400pt}}
\put(141.0,860.0){\rule[-0.200pt]{4.818pt}{0.400pt}}
\put(121,860){\makebox(0,0)[r]{20}}
\put(1419.0,860.0){\rule[-0.200pt]{4.818pt}{0.400pt}}
\put(141.0,123.0){\rule[-0.200pt]{0.400pt}{4.818pt}}
\put(141,82){\makebox(0,0){0}}
\put(141.0,840.0){\rule[-0.200pt]{0.400pt}{4.818pt}}
\put(401.0,123.0){\rule[-0.200pt]{0.400pt}{4.818pt}}
\put(401,82){\makebox(0,0){50}}
\put(401.0,840.0){\rule[-0.200pt]{0.400pt}{4.818pt}}
\put(660.0,123.0){\rule[-0.200pt]{0.400pt}{4.818pt}}
\put(660,82){\makebox(0,0){100}}
\put(660.0,840.0){\rule[-0.200pt]{0.400pt}{4.818pt}}
\put(920.0,123.0){\rule[-0.200pt]{0.400pt}{4.818pt}}
\put(920,82){\makebox(0,0){150}}
\put(920.0,840.0){\rule[-0.200pt]{0.400pt}{4.818pt}}
\put(1179.0,123.0){\rule[-0.200pt]{0.400pt}{4.818pt}}
\put(1179,82){\makebox(0,0){200}}
\put(1179.0,840.0){\rule[-0.200pt]{0.400pt}{4.818pt}}
\put(1439.0,123.0){\rule[-0.200pt]{0.400pt}{4.818pt}}
\put(1439,82){\makebox(0,0){250}}
\put(1439.0,840.0){\rule[-0.200pt]{0.400pt}{4.818pt}}
\put(141.0,123.0){\rule[-0.200pt]{312.688pt}{0.400pt}}
\put(1439.0,123.0){\rule[-0.200pt]{0.400pt}{177.543pt}}
\put(141.0,860.0){\rule[-0.200pt]{312.688pt}{0.400pt}}
\put(40,491){\makebox(0,0){Secs}}
\put(790,21){\makebox(0,0){Machines}}
\put(141.0,123.0){\rule[-0.200pt]{0.400pt}{177.543pt}}
\put(1279,246){\makebox(0,0)[r]{'chi\_file'}}
\put(1299.0,246.0){\rule[-0.200pt]{24.090pt}{0.400pt}}
\put(146,416){\usebox{\plotpoint}}
\multiput(146.58,416.00)(0.498,0.960){91}{\rule{0.120pt}{0.866pt}}
\multiput(145.17,416.00)(47.000,88.203){2}{\rule{0.400pt}{0.433pt}}
\multiput(193.58,506.00)(0.500,0.538){413}{\rule{0.120pt}{0.531pt}}
\multiput(192.17,506.00)(208.000,222.898){2}{\rule{0.400pt}{0.265pt}}
\multiput(401.00,730.58)(4.362,0.497){57}{\rule{3.553pt}{0.120pt}}
\multiput(401.00,729.17)(251.625,30.000){2}{\rule{1.777pt}{0.400pt}}
\multiput(660.00,760.58)(10.290,0.493){23}{\rule{8.100pt}{0.119pt}}
\multiput(660.00,759.17)(243.188,13.000){2}{\rule{4.050pt}{0.400pt}}
\multiput(920.00,773.58)(2.410,0.498){105}{\rule{2.019pt}{0.120pt}}
\multiput(920.00,772.17)(254.810,54.000){2}{\rule{1.009pt}{0.400pt}}
\multiput(1179.00,825.92)(5.414,-0.496){37}{\rule{4.360pt}{0.119pt}}
\multiput(1179.00,826.17)(203.951,-20.000){2}{\rule{2.180pt}{0.400pt}}
\put(146,416){\raisebox{-.8pt}{\makebox(0,0){$\Diamond$}}}
\put(193,506){\raisebox{-.8pt}{\makebox(0,0){$\Diamond$}}}
\put(401,730){\raisebox{-.8pt}{\makebox(0,0){$\Diamond$}}}
\put(660,760){\raisebox{-.8pt}{\makebox(0,0){$\Diamond$}}}
\put(920,773){\raisebox{-.8pt}{\makebox(0,0){$\Diamond$}}}
\put(1179,827){\raisebox{-.8pt}{\makebox(0,0){$\Diamond$}}}
\put(1392,807){\raisebox{-.8pt}{\makebox(0,0){$\Diamond$}}}
\put(1349,246){\raisebox{-.8pt}{\makebox(0,0){$\Diamond$}}}
\put(1279,205){\makebox(0,0)[r]{'ptcp'}}
\multiput(1299,205)(20.756,0.000){5}{\usebox{\plotpoint}}
\put(1399,205){\usebox{\plotpoint}}
\put(146,331){\usebox{\plotpoint}}
\multiput(146,331)(10.384,17.971){6}{\usebox{\plotpoint}}
\multiput(198,421)(19.009,8.334){10}{\usebox{\plotpoint}}
\multiput(401,510)(20.185,4.832){13}{\usebox{\plotpoint}}
\multiput(660,572)(20.207,4.741){13}{\usebox{\plotpoint}}
\multiput(920,633)(20.742,0.747){23}{\usebox{\plotpoint}}
\put(1392,650){\usebox{\plotpoint}}
\put(146,331){\makebox(0,0){$+$}}
\put(198,421){\makebox(0,0){$+$}}
\put(401,510){\makebox(0,0){$+$}}
\put(660,572){\makebox(0,0){$+$}}
\put(920,633){\makebox(0,0){$+$}}
\put(1392,650){\makebox(0,0){$+$}}
\put(1349,205){\makebox(0,0){$+$}}
\sbox{\plotpoint}{\rule[-0.400pt]{0.800pt}{0.800pt}}%
\put(1279,164){\makebox(0,0)[r]{'rcp'}}
\put(1299.0,164.0){\rule[-0.400pt]{24.090pt}{0.800pt}}
\put(146,202){\usebox{\plotpoint}}
\multiput(147.41,202.00)(0.505,7.305){35}{\rule{0.122pt}{11.476pt}}
\multiput(144.34,202.00)(21.000,272.181){2}{\rule{0.800pt}{5.738pt}}
\multiput(168.41,498.00)(0.504,7.777){41}{\rule{0.122pt}{12.267pt}}
\multiput(165.34,498.00)(24.000,336.540){2}{\rule{0.800pt}{6.133pt}}
\put(146,202){\raisebox{-.8pt}{\makebox(0,0){$\Box$}}}
\put(167,498){\raisebox{-.8pt}{\makebox(0,0){$\Box$}}}
\put(1349,164){\raisebox{-.8pt}{\makebox(0,0){$\Box$}}}
\end{picture}
      }
    \caption{Comparative Performance of ptcp}
    \label{fig:ptpcperf}
  \end{center}
\end{figure}
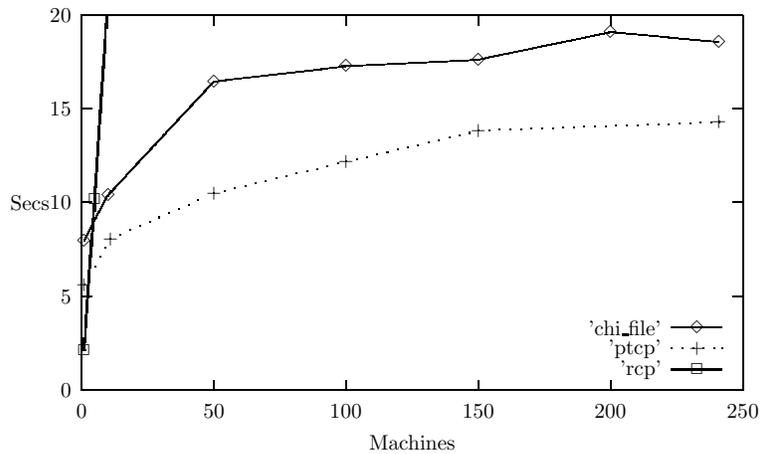

To justify the claims of scalability, we have carried out a small set of
experiments on Argonne's 256-node Chiba City cluster~\cite{chibacity}.
Execution times for simple commands are dominated by parallel process startup
time.  Commands that require substantial data movement are dominated by the
bandwidth of the communication links among the hosts and the algorithms used
to move data.  Timings for a trivial parallel task and one involving data
movement are shown in Table~\ref{tab:ptpcperf}.  Our copy test copies a 10MB
file that is randomly generated and does not compress well.  With text data
the effective bandwidth would be even higher.
\begin{table}[ht!]
\begin{center}
\caption{Performance of some commands}
\begin{tabular}{l|c|c|c|c|c|c}
\textbf{Number of Machines}&1&11&50&100&150&241\\
\hline
Time in seconds of a parallel&5.6&8.1&10.5&12.2&13.8&14.3\\
copy of 10MB over Fast Ethernet&&&&&&\\
\hline
Time in seconds of a parallel&0.8&0.9&1.2&1.5&1.8&1.9\\
execution of \texttt{hostname}&&&&&&\\
\end{tabular}
\label{tab:ptpcperf}
\end{center}
\end{table}
In Figure~\ref{fig:ptpcperf} we compare {\tt ptpc} with two other mechanisms
for copying a file to the local file systems on other nodes.  The simplest way
to do this is to call {\tt rcp} or {\tt scp} in a loop.  
Figure~\ref{fig:ptpcperf} shows how quickly this method becomes inferior to
more scalable approaches.  The ``chi\_file'' curve is for a sophisticated
system specifically developed for the Chiba City cluster~\cite{chibacity}.  This system,
written in Perl, takes advantage of the specific topology of the Chiba City
network and the way certain file systems are cross-mounted.  The general,
portable, MPI-based approach used by {\tt ptcp} performs better.

\section{Conclusion}
\label{sec:conclusion}
We have presented a design for an extension of the classical Unix tools to the
parallel domain, together with a scalable implementation using MPI.  The tools
are available at {\tt http://www.mcs.anl.gov/mpi}.  The distribution contains
all the necessary programs, complete source code, and {\tt man} pages for all
commands with much more detail than has been possible to present here.  An MPI
implementation is required; while any implementation should suffice, these
commands have been most extensively tested with
MPICH~\cite{gropp-lusk:mpich-www} and the MPD process
manager~\cite{bgl00:mpd:pvmmpi00}.  The tools are portable and can be 
installed on parallel machines running Linux, FreeBSD, Solaris, IRIX, or AIX.

\bibliography{/home/MPI/allbib}
\bibliographystyle{plain}

\end{document}